\documentclass [12pt] {article}
\usepackage{a4}

\usepackage{amsfonts}
\usepackage{amsmath}

  \def\pp{{\mathchoice
          {
              \kern 1pt%
              \raise 1pt
              \vbox{\hrule width5pt height0.4pt depth0pt
                    \kern -2pt
                    \hbox{\kern 2.3pt
                          \vrule width0.4pt height6pt depth0pt
                          }
                    \kern -2pt
                    \hrule width5pt height0.4pt depth0pt}%
                    \kern 1pt
           }
            {
              \kern 1pt%
              \raise 1pt
              \vbox{\hrule width4.3pt height0.4pt depth0pt
                    \kern -1.8pt
                    \hbox{\kern 1.95pt
                          \vrule width0.4pt height5.4pt depth0pt
                          }
                    \kern -1.8pt
                    \hrule width4.3pt height0.4pt depth0pt}%
                    \kern 1pt
            }
            {
              \kern 0.5pt%
              \raise 1pt
              \vbox{\hrule width4.0pt height0.3pt depth0pt
                    \kern -1.9pt  
                    \hbox{\kern 1.85pt
                          \vrule width0.3pt height5.7pt depth0pt
                          }
                    \kern -1.9pt
                    \hrule width4.0pt height0.3pt depth0pt}%
                    \kern 0.5pt
            }
            {
              \kern 0.5pt%
              \raise 1pt
              \vbox{\hrule width3.6pt height0.3pt depth0pt
                    \kern -1.5pt
                    \hbox{\kern 1.65pt
                          \vrule width0.3pt height4.5pt depth0pt
                          }
                    \kern -1.5pt
                    \hrule width3.6pt height0.3pt depth0pt}%
                    \kern 0.5pt
            }
        }}

  \def\mm{{\mathchoice
                       {
                             \kern 1pt
               \raise 1pt    \vbox{\hrule width5pt height0.4pt depth0pt
                                  \kern 2pt
                                  \hrule width5pt height0.4pt depth0pt}
                             \kern 1pt}
                       {
                            \kern 1pt
               \raise 1pt \vbox{\hrule width4.3pt height0.4pt depth0pt
                                  \kern 1.8pt
                                  \hrule width4.3pt height0.4pt depth0pt}
                             \kern 1pt}
                       {
                            \kern 0.5pt
               \raise 1pt
                            \vbox{\hrule width4.0pt height0.3pt depth0pt
                                  \kern 1.9pt
                                  \hrule width4.0pt height0.3pt depth0pt}
                            \kern 1pt}
                       {
                           \kern 0.5pt
             \raise 1pt  \vbox{\hrule width3.6pt height0.3pt depth0pt
                                  \kern 1.5pt
                                  \hrule width3.6pt height0.3pt depth0pt}
                           \kern 0.5pt}
                       }}

\catcode`@=11
\def\un#1{\relax\ifmmode\@@underline#1\else
        $\@@underline{\hbox{#1}}$\relax\fi}
\catcode`@=12

\let\du=\du                     

\def\a{\alpha}
\def\b{\beta}

\def\d{\delta}

\def\f{\phi}
\def\g{\gamma}
\def\h{\eta}

\def\j{\psi}

\def\m{\mu}
\def\n{\nu}
\def\o{\omega}
\def\p{\pi}
\def\q{\theta}
\def\r{\rho}
\def\s{\sigma}
\def\t{\tau}

\def\z{\zeta}
\def\D{\Delta}

\def\G{\Gamma}

\def\L{\Lambda}

\def\Q{\Theta}

\def\X{\Xi}

\def\ve{\varepsilon}

\def\cd{{\cal D}}

\def\ch{{\cal H}}

\def\cw{{\cal W}}

\def\bo{{\raise-.3ex\hbox{\large$\Box$}}}               
\def\pa{\partial}                                       
\def\TH{{\raise.2ex\hbox{$\displaystyle \bigodot$}\mskip-4.7mu \llap H \;}}
\def\face{{\raise.2ex\hbox{$\displaystyle \bigodot$}\mskip-2.2mu \llap {$\ddot
        \smile$}}}                                      

\def\abs#1{\left| #1\right|}                    
\def\leftrightarrowfill{$\mathsurround=0pt \mathord\leftarrow \mkern-6mu
        \cleaders\hbox{$\mkern-2mu \mathord- \mkern-2mu$}\hfill
        \mkern-6mu \mathord\rightarrow$}
\def\dvec#1{\vbox{\ialign{##\crcr
        \leftrightarrowfill\crcr\noalign{\kern-1pt\nointerlineskip}
        $\hfil\displaystyle{#1}\hfil$\crcr}}}           

\def\frac#1#2{{\textstyle{#1\over\vphantom2\smash{\raise.20ex
        \hbox{$\scriptstyle{#2}$}}}}}                   
\def\sfrac#1#2{{\vphantom1\smash{\lower.5ex\hbox{\small$#1$}}\over
        \vphantom1\smash{\raise.4ex\hbox{\small$#2$}}}} 
\def\bfrac#1#2{{\vphantom1\smash{\lower.5ex\hbox{$#1$}}\over
        \vphantom1\smash{\raise.3ex\hbox{$#2$}}}}       
\def\afrac#1#2{{\vphantom1\smash{\lower.5ex\hbox{$#1$}}\over#2}}    

\def\[{\lfloor{\hskip 0.35pt}\!\!\!\lceil}
\def\]{\rfloor{\hskip 0.35pt}\!\!\!\rceil}
\def\Lag{{\cal L}}
\def\du#1#2{_{#1}{}^{#2}}

\def\fracm#1#2{\hbox{\large{${\frac{{#1}}{{#2}}}$}}}

\def\tr{{\rm tr}}

\def\un{\underline}
\def\fracmm#1#2{{{#1}\over{#2}}}

\def\low#1{{\raise -3pt\hbox{${\hskip 0.75pt}\!_{#1}$}}}

\newskip\humongous \humongous=0pt plus 1000pt minus 1000pt
\def\caja{\mathsurround=0pt}
\def\eqalign#1{\,\vcenter{\openup2\jot \caja
        \ialign{\strut \hfil$\displaystyle{##}$&$
        \displaystyle{{}##}$\hfil\crcr#1\crcr}}\,}
\newif\ifdtup

\def\pl#1#2#3{Phys.~Lett.~{\bf {#1}B} (19{#2}) #3}
\def\np#1#2#3{Nucl.~Phys.~{\bf B{#1}} (19{#2}) #3}

\def\cqg#1#2#3{Class.~and Quantum Grav.~{\bf {#1}} (19{#2}) #3}
\def\cmp#1#2#3{Commun.~Math.~Phys.~{\bf {#1}} (19{#2}) #3}

\parskip=\medskipamount                 
\thicklines                         

\begin{document}
\thispagestyle{empty}

{\hbox to\hsize{
\vbox{\noindent December 2002   \hfill hep-th/0212003 }}}

\noindent
\vskip1.3cm
\begin{center}

{\Large\bf Instanton-induced scalar potential \vglue.2in
            for the universal hypermultiplet}

\vglue.2in

Sergei V. Ketov 

{\it Department of Physics\\
     Tokyo Metropolitan University\\
     1--1 Minami-osawa, Hachioji-shi\\
     Tokyo 192--0397, Japan}\\
{\sl ketov@comp.metro-u.ac.jp}
\end{center}
\vglue.2in
\begin{center}
{\Large\bf Abstract}
\end{center}

\noindent We calculate the scalar potential in the gauged N=2 supergravity with
a single hypermultiplet, whose generic quaternionic moduli space metric has an
abelian isometry. This isometry is gauged by the use of a graviphoton gauge 
field. The hypermultiplet metric and the scalar potential are both governed by
the single real potential that is a solution to the 3d (integrable) continuous
Toda equation. An explicit solution, controlled by the Eisenstein series 
$E_{3/2}$, is found in the case of the D-instanton-corrected universal 
hypermultilet moduli space metric having an $U(1)\times U(1)$ isometry, with 
one of the isometries being gauged.  

\newpage

\section{Introduction}

The {\it Universal Hypermultiplet} (UH) sector of the {\it Calabi-Yau} (CY)  
compactified type-IIA superstrings/M-theory is a good place to study 
non-perturbative quantum corrections within the effective N=2 supergravity 
in four or five spacetime dimensions \cite{bbs}. The UH contains a dilaton 
$\f$, an axion $D$ and a RR-type complex scalar $C$ as the bosonic field 
components, while the UH is present in {\it any} CY compactification of 
type-IIA superstrings. The classical UH  moduli space is given by a symmetric 
(homogeneous)  quaternionic space $SU(2,1)/U(2)$ \cite{fsh}, while its metric 
and isometries are not protected against quantum corrections on the type-IIA 
side. The perturbative (type-IIA superstring loop) corrections to the UH 
metric are known to be limited to the one-loop order, being proportional to 
the Euler characteristics of CY \cite{one}. The origin of the non-perturbative 
corrections is also well understood \cite{bbs}: they appear due to the 
so-called D-instantons and five-brane instantons \cite{w1}. The former are the
 Euclidean D2-branes wrapped about the supersymmetric 3-cycles of CY, whereas 
the latter are the Euclidean BPS five-branes wrapped about the entire CY space
 \cite{bbs}. The relevant instanton solutions saturating the BPS bound, as 
well as the corresponding instanton actions, were calculated in 
ref.~\cite{actions}.

The quantum UH moduli space metric is highly constrained by unbroken 
symmetries. This metric must be quaternionic because of unbroken N=2 local 
supersymmetry in four or five uncompactified spacetime dimensions \cite{bw}.
 As regards a single hypermultiplet (like UH) with the {\it four}-dimensional 
moduli space, the quaternionic condition amounts to the {\it Einstein-Weyl\/} 
equations (see sect.~3 for details). The quantized brane charges (or the flux 
quantization condition of the antisymmetric tensor field in M-theory) imply 
discrete identifications for the UH scalars, which break most of the 
continuous classical symmetries of $SU(2,1)$. Nevertheless, the $U(1)$ 
rotations of the RR-scalar, $C\to e^{i\a}C$, survive after taking into 
account the instanton corrections. The extra abelian symmetry associated with
 constant shifts of the axion, $D\to D+\d$, also survives when merely 
D-instantons are taken into account and the five-brane instantons are 
suppressed \cite{ket}. These observations are consistent with the known
instanton actions \cite{actions}. In particular, the exact UH metric is 
governed by the single pre-potential that is a solution to the 
three-dimensional (integrable) Toda equation \cite{ket}. The D-instanton 
corrected quantum moduli space metric of the UH is supposed to be 
$SL(2,{\bf Z})$-duality invariant. Its explicit form was found in 
ref.~\cite{ket}, in terms of the $E_{3/2}$ Eisenstein series, in agreement with
 the supersymmetric completion of the $R^4$-terms in ten-dimensional 
superstrings \cite{green}. Some explicit results about the five-brane 
instanton corrected UM moduli space metric were obtained in ref.~\cite{nbi},
in terms of the particular exact solution to the Painlev\'e VI (integrable)
equation \cite{pen}.

An addition of non-trivial fluxes of the NS-NS and R-R three-forms in 
ten dimensions amounts to {\it gauging} some Peccei-Quinn-type isometries 
of the UH moduli space in the effective N=2 supergravity \cite{ps}. As a 
result of the gauging, the UH gets the non-trivial {\it scalar} potential 
whose critical points determine the vacua of the theory \cite{ps}. An explicit 
gauging of all abelian isometries of the {\it classical} UH moduli space 
metric was performed in ref.~\cite{clg}. As regards the {\it quantum} UH 
metric, one can merely gauge the single abelian $U(1)$ isometry that survives 
after adding quantum instanton corrections. Gauging the abelian isometries of 
the classical UH moduli space metric gives rise to the scalar potential with 
the unphysical run-away behaviour, or no critical points in the weak-coupling 
region where perturbation theory applies \cite{clg}. Since the classical UH 
scalar potential is not protected against quantum corrections, it is more
physically reasonable to examine the instanton-corrected UH scalar potential.

Another important motivation to study gauging of an abelian 
isometry of the instanton-corrected hypermultiplet metric is its relevance to 
the brane-world scenario \cite{bworld}  in the effective five-dimensional N=2
 gauged supergravity.~\footnote{There is no difference  in treating 
hypermultiplets in N=2 supergravities in four and five spacetime \newline
${~~~~~}$ dimensions.}  The brane world scenario (with gravity trapped near a 
domain wall) needs a scalar potential with at least two IR critical points, in
 order to achieve an exponential suppression on both sides of the wall.
 In the context of the gauged N=2 supergravity, this can only
 be achieved by gauging an isometry of a non-homogeneous hypermultiplet moduli 
space \cite{prob}. Some explicit examples of such construction were given in 
refs.~\cite{toy,sbl}. Unfortunately, the hypermultiplet metrics used in 
refs.~\cite{toy,sbl} were chosen {\it ad hoc}, they have unphysical regions, 
and they were not derived from some underlying theory (like superstings or 
M-theory). The instanton-corrected hypermultiplet moduli space is not 
homogeneous, while it also gives the natural physical input towards a possible
brane world scenario in the CY compactified type-IIA superstrings or M-theory. 

Our paper is organized as follows. In sect.~2 we review the relevant facts 
about gauging an abelian isometry of a hypermultiplet metric in N=2 
supergravity. Our discussion is limited to the scalar potential of a single 
hypermultiplet. In sect.~3 we discuss the relation between the Einstein-Weyl 
spaces with an $U(1)$ abelian isometry and integrable systems. An explicit
solution to the D-instanton-induced UH scalar potential is given in sect.~4.
The critical points of the scalar potentials are discussed in sect.~5. Sect.~6
is our conclusion. A brief review of the Einstein-Weyl geometry and a summary 
of our notation are given in Appendix.

\section{Hypermultiplet scalar potential in the gauged N=2 supergravity}

Our purpose in this section is to provide {\it minimum} information needed to
calculate a scalar potential in the gauged N=2 supergravity with a {\it single}
charged hypermultiplet whose quaternionic metric has an abelian isometry. This
will serve as the pre-requisite for the subsequent sections.

The detailed structure of a generic gauged N=2 supergravity theory with a
hypermultiplet matter in four or five spacetime dimensions is well known 
(see, e.g. ref.~\cite{fer} for a recent account). In our case, all the
relevant formulae can be extracted from the most recent paper \cite{toy} that 
we are going to follow in this section. 

The field contents of an N=2 supergravity multiplet is given by a graviton 
$e^a_{\m}$, two Majorana gravitinos $\j^{\a}_{\m i}$, $i=1,2$, and a 
graviphoton (an abelian vector gauge field) $A_{\m}$. The field contents of a 
hypermultiplet is given by four real hyperscalars $q^X$ and a Dirac hyperino 
$\h_{\a}$ (for definiteness, we refer to four spacetime dimensions, 
$\m=0,1,2,3$, etc.).~\footnote{The 4-dimensional (curved) spacetime should 
not be confused with the 4-dimensional (curved) \newline ${~~~~~}$  
hypermultiplet moduli space. We also distinguish between the hypermultiplet 
moduli space and \newline ${~~~~~}$ its tangent space, as well as between the
corresponding indices --- see Appendix.}
 In the case of the UH, the scalars $q^X$ represent a dilaton, an axion and a 
complex RR-type scalar in an arbitrary (non-linear sigma-model) 
parametrization.

The relevant bosonic part of the hypermultiplet low-energy effective action in
 N=2 supergravity is given by 
$$ e^{-1}\Lag = -\fracmm{1}{2}R -\fracmm{1}{4}F_{\m\n}F^{\m\n}
-\,\fracmm{1}{2}g\low{XY}\cd_{\m}q^X\cd^{\m}q^Y -g^2V~,\eqno(2.1)$$
where we have added the standard kinetic terms for the graviton 
(Einstein-Hilbert) and the graviphoton (Maxwell). The kinetic terms of the 
hypermultiplet in eq.~(2.1) are given by the gauged 
{\it Non-Linear Sigma-Model} (NLSM) with the four-dimensional quaternionic 
metric $g\low{XY}(q)$, and the gauge-covariant derivatives
$$ \cd_{\m}q^X =\pa_{\m}q^X +gA_{\m}k^X(q)~,\eqno(2.2)$$
in terms of the Killing vector $k^X(q)$ of the gauged isometry of the
hypermultiplet moduli space parameterized by $q^X$, and the gauge coupling
constant $g$. The scalar potential $V(q)$ in eq.~(2.1) is given by \cite{ps}
$$ V = -4P^kP^k +\fracmm{3}{4}g\low{XY}k^Xk^Y~,\eqno(2.3)$$
where a triplet of the Killing pre-potentials $P^k$ has been introduced,
$$ P^k= -\fracmm{1}{4} D_Xk_Y J^{XYk}= -\fracmm{1}{4}\pa_Xk_YJ^{XYk}~,\quad 
k=1,2,3~,\eqno(2.4)$$
in terms of the complex structures $J^{XYk}=-J^{YXk}$ of the quaternionic 
metric $g\low{XY}$. However, eq.~(2.3) is not convenient for our purposes, 
since it requires a calculation of the quaternionic structure $J^{XYk}$ that 
is not really needed.

In fact, the structure of the scalar potential $V$ in eq.~(2.1) is dictated by
another scalar function known as the {\it superpotential} $W$ that can be read
off from the gravitino supersymmetry transformation law \cite{fer},
$$ \d\j_{\m i} =D_{\m}\ve_i +\fracmm{ig}{\sqrt{6}}\g_{\m}P_{ij}\ve^j+\ldots~,
\quad i,j=1,2~,\eqno(2.5)$$
where merely the bosonic terms have been written down on the right-hand-side.
The superpotential $W$ is defined in terms of the Killing pre-potentials
 \cite{fer},
$$ W^2 = \fracmm{1}{3}P_{ij}P^{ij}= \fracmm{2}{3}P^{k}P^{k}~,\eqno(2.6)$$
whereas the scalar potential $V$ is related to the superpotential $W$ as
follows \cite{toy}:
$$ V = -6W^2 +\fracmm{9}{2} g^{XY}\pa_XW\pa_YW~~.\eqno(2.7)$$
In the special case of a single hypermultiplet, the quaternionic identity
$J^k\low{XY}J^k\low{ZW}=-\ve\low{XYZW}
+(\d\low{XZ}\d\low{YW}-\d\low{XW}\d\low{YZ})$ allows one to 
rewrite eq.~(2.6) into a simpler form \cite{toy}
$$ W^2 =\fracmm{1}{3}dK\wedge {}^*dK -\fracmm{1}{6}dK\wedge dK~,\eqno(2.8)$$
in terms of the Killing one-form $K=k_Xdq^X$ and the Hodge star operation 
$(*)$ alone.

Thus the only NLSM reparametrization-invariant input needed to calculate the 
scalar potential $V$ is given by a quaternionic metric $g\low{XY}$ and its 
Killing vector $K$, by using eqs.~(2.7) and (2.8). In physical applications we
should choose a particular parametrization of the hypermultiplet moduli 
space, in which the abelian isometry of the metric is manifest (see sect.~3).

It is worth noticing that the scalar potential $V(q)$ is obviously dependent 
upon the chosen NLSM parametrization of the hypermultiplet scalars, whereas the
critical points (vacua) of the scalar potential are 
parametrization-independent. Under a reparametrization $q=q(\tilde{q})$ we have
$$ \fracmm{\pa V}{\pa\tilde{q}}=
\fracmm{\pa q}{\pa\tilde{q}}\fracmm{\pa V}{\pa q}~,\eqno(2.9)$$
so that $\pa V/\pa q=0$ is equivalent to  $\pa V/\pa\tilde{q}=0$ because of 
$\det(\pa q/\pa\tilde{q})\neq 0$. 

\section{Einstein-Weyl metrics with an abelian isometry}

An N=2 locally supersymmetric NLSM with any number of hypermultiplets has a 
quaternionic metric \cite{bw}. In the case of a single hypermultiplet, N=2
local supersymmetry amounts to the Einstein-Weyl conditions (with a negative
scalar curvature) on the NLSM metric $g$ \cite{bw},~\footnote{Our notation is 
given in Appendix.}
$$ W^-_i=0~,\quad (Ric)_{ab}=\L\d_{ab}~,\quad \L<0~.\eqno(3.1)$$

Given an isometry of the metric $g$ with the associated Killing 1-form 
$K=K_Xdq^X$ and $D_XK_Y+D_YK_X=0$, one can decompose the 
2-form $dK=\pa_XK_Ydq^X\wedge dq^Y$ with respect to the basis 
(A.8),
$$ dK= (dK^+_i)\X^+_i + (dK^-_i)\X^-_i~,\eqno(3.2)$$
and extract the quaternionic structure (i.e. three complex structures $J^k$ 
obeying the quaternionic algebra $J^iJ^j=-\d^{ij}+\ve^{ijk}J^k$) in terms of 
the Killing form \cite{prz,tod,iva},
$$ J = \fracmm{dK^-_i}{\sqrt{\sum_i (dK^-_i)^2}}\,\X^-_i~~.\eqno(3.3)$$

An important theorem due to Przanowski \cite{prz}  and Tod \cite{tod} claims 
that any Einstein-Weyl metric can be locally written down in adapted 
coordinates (with a Killing vector $\pa_t$) as follows:
$$ g= \fracmm{1}{w^2}\left\{ \fracmm{1}{P}(dt+\Q)^2+P\left[e^u(d\m^2+d\n^2)+
dw^2\right]\right\}~.\eqno(3.4)$$
in terms of real local coordinates $(t,w,\m,\n)$, 1-form 
$\Q=\Q_1dw+\Q_2d\m+\Q_3d\n$, and two potentials $P=P(w,\m,\n)$ and
$u=u(w,\m,\n)$. Imposing the Einstein-Weyl conditions (3.1) on the metric (3.4)
yields \cite{prz,tod}
$$ P = \fracmm{3}{2\L}\left(w\pa_wu -2 \right)~,\eqno(3.5)$$
$$ \left(\pa^2_{\m}+\pa^2_{\n}\right)u +\pa^2_w(e^u)=0~,\eqno(3.6)$$
and
$$ -d\Q=(\pa_{\n}P)d\m\wedge dw+(\pa_{\m}P)dw\wedge d\n+\pa_w(Pe^u)d\n\wedge
d\m~.\eqno(3.7)$$
As is clear from eqs.~(3.5), (3.6) and (3.7), the metric $g$ is controlled by
the single pre-potential $u$ obeying the non-linear (integrable) 
three-dimensional (continuous) Toda equation (3.6).

In accordance to eq.~(3.4), let's choose the vierbein as
$$ \eqalign{
e_0= \fracmm{(dt+\Q)}{w\sqrt{P}}~,\quad & \quad 
e_2= \sqrt{P}\exp(\frac{1}{2}u)\fracmm{d\m}{w}~,\cr
e_1= \sqrt{P}\,\fracmm{dw}{w}~,\quad & \quad 
e_3= \sqrt{P}\exp(\frac{1}{2}u)\fracmm{d\n}{w}~.\cr}\eqno(3.8)$$

The Killing vector $K^X=(1,0,0,0)$ yields the Killing 1-form
$$ K= \fracmm{1}{w^2P}(dt+\Q)=\fracmm{1}{w\sqrt{P}}\,e_0~.\eqno(3.9)$$
The square of the Killing vector is given by
$$ K^2= g_{XY}K^XK^Y=g_{tt}=\fracmm{1}{w^2P}~~.\eqno(3.10)$$
The coordinate $w$ in terms of the Killing vector $K$ reads \cite{iva}
$$ w= \fracmm{-\L/3}{\sqrt{\sum_i (dK^-_i)^2}}~~.\eqno(3.11)$$

By using the identities
$$ \X^+_i\wedge \X^+_j=- \X^-_i\wedge \X^-_j=2\d_{ij}e_0\wedge e_1\wedge e_2
\wedge e_3 \eqno(3.12a)$$
and
$$ \X^+_i\wedge \X^-_j=\X^-_i\wedge \X^+_j=0~,\eqno(3.12b)$$
and substituting the decomposition (3.2) into eq.~(2.8) allows us to simplify
the superpotential $W$ to the form
$$ W^2= (dK^-_i)^2+\frac{1}{3}(dK^+_i)^2=\fracmm{\L^2}{9w^2} 
+\frac{1}{3}(dK^+_i)^2~,\eqno(3.13)$$
where eq.~(3.11) has been used.

It is straightforward to calculate the 2-form $dK$ from eqs.~(3.8) and (3.9).
We find
$$ \eqalign{
dK =~ & \fracmm{1}{\sqrt{P}}\left( 1 + \fracmm{w}{2P}\pa_wP\right)e_0\wedge e_1
\cr
& +\fracmm{w\pa_{\m}P}{2P\sqrt{e^uP}}e_0\wedge e_2
+ \fracmm{w\pa_{\n}P}{2P\sqrt{e^uP}}e_0\wedge e_3 \cr
& +  \fracmm{w\pa_{\n}P}{P\sqrt{e^uP}}e_1\wedge e_2
+\fracmm{w\pa_{\m}P}{P\sqrt{e^uP}}e_3\wedge e_1 \cr
& + \fracmm{w}{\sqrt{P}}\left(\pa_wu+\fracmm{\pa_wP}{P}\right)e_2\wedge e_3~.}
\eqno(3.14)$$

Equations (3.2) and (A.8) now imply
$$\eqalign{
(dK^+_i)\X^+_i=~ & \fracmm{1}{2\sqrt{P}}\left(1+ \fracmm{3w}{2P}\pa_wP 
+w\pa_wu\right)\X^+_1 \cr
& + \fracmm{3w\pa_{\m}P}{2P\sqrt{e^uP}}\,\X^+_2 + 
\fracmm{3w\pa_{\n}P}{2P\sqrt{e^uP}}\,\X^+_3~,\cr}\eqno(3.15)$$
and hence, we get
$$ (dK^+_i)^2=\fracmm{1}{4P}\left(1+w\pa_wu+\fracmm{3}{2P}w\pa_wP\right)^2
+\fracmm{9w^2e^{-u}}{4P^3}\left[ (\pa_{\m}P)^2 + (\pa_{\n}P)^2\right]~.
\eqno(3.16)$$ 

We conclude that both the hypermultiplet metric (3.4) {\it and} the scalar 
potential (2.7) are dictated by a solution $u(w,\m,\n)$ of the Toda equation 
(3.6) via eqs.~(3.5), (3.7), and eqs.~(2.7), (3.13) and (3.16), respectively.

By substituting eq.~(3.16) into eq.~(3.13) and using eq.~(3.5), we find the
superpotential in the form
$$\eqalign{
W^2=~ &  \fracmm{\L^2}{9w^2}+\fracmm{1}{12P}\left(3+\fracmm{2\L}{3}P+
\fracmm{3}{2P}w\pa_wP\right)^2 \cr
 & +\fracmm{3w^2e^{-u}}{4P^3}\left[ (\pa_{\m}P)^2+ (\pa_{\n}P)^2\right]~.\cr}
\eqno(3.17)$$ 

Until this point no approximation was made, so that we actually discussed a
derivation of {\it exact} solutions to the hypermultiplet moduli space metric 
and the scalar potential. Unfortunately, despite of the fact that the Toda 
equation (3.6) is known to be integrable (this equation appears in the
large-$N$ limit of the standard (two-dimensional) Toda system for $SU(N)$ 
\cite{ward}), it is very hard to obtain its explicit solutions \cite{lez}. This
is apparently the price to pay for getting the exact solution describing 
both five-brane and two-brane instanton corrections to the UH metric and its 
scalar potential \cite{nbi}.

It is instructive to see how this problem simplifies in the 
{\it hyper-K\"ahler} limit for the hypermultiplet metric, when N=2 
supergravity decouples. This limit appears when $\L\to 0$ above, since $\L$ is
 proportional to the gravitational coupling constant \cite{bw}. In this limit 
the function $P$ becomes proportional to $\pa_w u$, whereas the non-linear 
Toda equation (3.6) becomes a {\it linear} equation on $P$ \cite{lebrun,nbi},
$$ (\pa^2_{\m}+\pa^2_{\n}+\pa^2_{w})P=0~.\eqno(3.18)$$
The abelian isometry is tri-holomorphic in this limit, so that we obtain the
standard Gibbons-Hawking {\it Ansatz} for a hyper-K\"ahler metric governed by
a harmonic function $P(w,\m,\n)$ \cite{gh}. 

The scalar potential, originating from the gauging of the tri-holomorphic
isometry in the hyper-K\"ahler limit is given by half of the Killing vector 
squared, or just $\frac{1}{2}P^{-1}$. For example, the (Gibbons-Hawking) 
multi-centre metrics are described by
$$ P(\vec{X}) =\sum^m_{p=1} \fracmm{1}{{\abs{\vec{X}-\vec{X}_p}}} ~~~,\qquad
\vec{X}=(w,\m,\n)~,\quad \vec{X}_p=const~.\eqno(3.19)$$
The critical points of the scalar potential are given by {\it poles} of $P$, 
i.e. they appear at $\vec{X}=\vec{X}_p$ in the case of eq.~(3.19). Since the 
scalar potential vanishes at these points, N=2 supersymmetry remains unbroken.
 Our results are, therefore, consistent with a derivation of the 
hypermultiplet scalar potential by Scherk-Schwarz dimensional reduction from 
six dimensions in the hyper-K\"ahler limit \cite{tong}.

\section{UH scalar potential induced by D-instantons}
 
To get an explicit non-perturbative solution to the hypermultiplet scalar 
potential, we now consider the special case of the UH when the D-instanton 
contributions are included but the five-brane instantons are suppressed. The
D-instanton corrections are of the order $e^{-1/g_{\rm string}}$, whereas the
five-brane instanton corrections are of the order $e^{-1/g^2_{\rm string}}$
\cite{w1}. Hence, for sufficiently small string coupling $g_{\rm string}$, we
may hope that the D-instanton corrections dominate over the five-brane
instanton corrections. In this case, there is another abelian isometry given 
by a shift of the axion, $D\to D+\d$, which commutes with an $U(1)$ rotation of
the RR-scalar, $C\to e^{i\a}C$, that is going to be gauged. As was pointed out
in the second ref.~\cite{clg}, gauging a compact direction of the homogeneous
hypermultiplet moduli space yields a fixed point, whereas gauging a non-compact
direction yields a run-away solution.

Due to some recent advances in the mathematical literature \cite{cp}, given
 two commuting and non-degenerate (i.e. hypersurface generating) abelian 
isometries (Killing vectors), one can completely solve the Einstein-Weyl
equations (3.1) in adapted coordinates, where {\it both} isometries are 
manifest, in terms of a real potential depending upon two remaining 
coordinates and satisfying a {\it linear} equation. 

The main result of ref.~\cite{cp} is the theorem that {\it any} Einstein-Weyl
metric (of non-vanishing scalar curvature) with two linearly independent 
Killing vectors can be written down in the from 
$$\eqalign{ 
 g ~=~ &  \fracmm{F^2-4\r^2(F^2_{\r}+F^2_{\h})}{4F^2}\,
\left(\fracmm{d\r^2+d\h^2}{\r^2}\right) \cr 
 & + \fracmm{ [(F-2\r F_{\r})\hat{\a}-2\r F_{\h}\hat{\b} ]^2 +[2\r F_{\h}
\hat{\a}-(F+2\r F_{\r})\hat{\b}]^2 }{F^2[F^2-4\r^2(F^2_{\r}+F^2_{\h})] }~,\cr}
\eqno(4.1)$$
in some local coordinates $(\r,\h,\q,t)$ inside an open region of the 
half-space $\r>0$. Here $\pa_{\q}$ and $\pa_{t}$ are two Killing 
vectors, while the one-forms $\hat{\a}$ and $\hat{\b}$ are given by
$$ \hat{\a}= \sqrt{\r}\,d\q\quad {\rm and}\quad \hat{\b}=\fracmm{dt 
+\h d\q}{\sqrt{\r}}~~.\eqno(4.2)$$

The whole metric (4.1) is governed by a real function (= {\it pre-potential}) 
$F(\r,\h)$ that is the eigenfunction of the Laplacian in the hyperbolic plane,
$$\D_{\ch}F \equiv \r^2\left(\pa^2_{\r}+\pa^2_{\h}\right)F =
\fracmm{3}{4}F~~.\eqno(4.3)$$

The Einstein-Weyl metric (4.1) has a negative scalar curvature provided that
$$ 4\r^2(F^2_{\r}+F^2_{\h}) > F^2>0~~.\eqno(4.4)$$

As was demonstrated in ref.~\cite{ket}, an unique (up to a normalization) 
$SL(2,{\bf Z})$ duality-invariant solution to the master equation (4.3) is 
given by the {\it Eisenstein} series $E_{3/2}(\r,\h)$. It has the Fourier 
expansion \cite{ter}
$$ 4\p\z(3) E_{3/2}(\r,\h)= 2\z(3) \r^{3/2} +\fracmm{2\p^2}{3}
\r^{-1/2} + 8\p\r^{1/2}\sum_{m\neq 0 \atop n\geq 1} \abs{\fracmm{m}{n}}
e^{2\p imn\h}K_1(2\p\abs{mn}\r)~,\eqno(4.5)$$
where $\z(3)= \sum_{m>0}(1/m)^3$ and the modified Bessel function $K_1(z)$ of
the 3rd kind have been introduced. The asymptotic expansion of the 
hypermultiplet pre-potential in the perturbative region (large $\r$) reads
$$ F(\r,\h)=  4\p\z(3) E_{3/2}(\r,\h) = 2\z(3)\r^{3/2} 
+\fracmm{2\p^2}{3}\r^{-1/2} +4\p^{3/2}\sum_{m,n\geq 1}\left(\fracmm{m}{n^3}
\right)^{1/2}\times$$
$$\times \left[ e^{2\p i mn(\h+i\r)} + e^{-2\p i mn(\h-i\r)}\right] 
\left[ 1 + \sum^{\infty}_{k=1}\fracmm{\G(k-1/2)}{\G(-k-1/2)}\,
\fracmm{1}{(4\p mn\r)^k}\right]~,\eqno(4.6)$$
while it can be interpreted as a sum of the classical (tree level) term, the
one-loop (perturbative) correction and the infinite D-instanton sum, 
respectively, in the apparent similarity to the known $SL(2,{\bf Z})$ 
duality-invariant completion of the $R^4$-terms in the ten-dimensional type-IIB
 superstrings \cite{green}. We expect that our result (4.6) can be 
reproduced from the ten-dimensional $R^4$-terms via CY compactification 
\cite{ket}.

It is not difficult to map the Calderbank-Petersen (=CP) {\it Ansatz} (4.1) 
into the more general Przanowski-Tod (=PT) {\it Ansatz} (3.4). In fact, this 
was already done in ref.~\cite{iva}. We are going to pay a special attention 
to the PT coordinate $w$ and the PT potential $P$ in terms of the `active' CP 
coordinates $(\r,\h)$ and the CP pre-potential $F(\r,\h)$. In terms of the
related function 
$$ G =\sqrt{\r}F~,\eqno(4.6)$$
the CP metric (4.1) can be rewritten to the form \cite{iva}
$$ g = \fracmm{1}{G^2}\left\{ \fracmm{1}{\cw}(dt+\Q)^2 +\cw\g\right\}~,
\eqno(4.7)$$
where we have used the notation \cite{iva}
$$ \cw=\fracmm{GG_{\r}}{\r(G^2_{\r}+G^2_{\h})}-1~,\qquad
\Q= \left( \fracmm{GG_{\h}}{G^2_{\r}+G^2_{\h}}-\h\right)d\a~,\eqno(4.8a)$$
and
$$ \g = \r^2d\a^2 +(G^2_{\r}+G^2_{\h})(d\r^2+d\h^2)~~.\eqno(4.8b)$$

Let $K$ be the 1-form associated with the Killing vector $\pa_t$,
$$ K = \fracmm{dt+\Q}{G^2\cw}~.\eqno(4.9)$$
We can now explicitly compute the 2-form $dK$, as well as its SD and ASD 
parts, $dK^-$ and $dK^+$, like in the previous sect.~3. For example, one 
finds \cite{iva}
$$ dK^-= \fracmm{-1}{G\sqrt{G^2_{\r}+G^2_{\h}}}\left(G_{\r}\X^-_1+
G_{\h}\X^-_2\right)~.\eqno(4.10)$$
This allows us to identify 
$$ w=G \quad{\rm and}\quad P=\cw~~.\eqno(4.11)$$
More explicitly, we find
$$ w = \sqrt{\r}\,F = \sqrt{\r}\,E_{3/2}(\r,\h)~,\eqno(4.12)$$
and
$$P = \fracmm{E^2_{3/2}+\r\pa E^2_{3/2}/\pa\r}{2\r^2\left[
\left( \fracmm{E_{3/2}}{2\r}+\fracmm{\pa E_{3/2}}{\pa\r}\right)^2+
\left(\fracmm{\pa E_{3/2}}{\pa\h}\right)^2\right]}~~-1~.\eqno(4.13)$$

Once the $P$-function is known, the Toda potential $u$ is easily obtained by
integrating eq.~(3.5). The final result for the D-instanton-corrected scalar 
potential (or the superpotential) of the UH, in terms of the Eisenstein series 
$E_{3/2}$, is not very illuminating, so that we do not write it down here. 
Instead, in the next sect.~5, we discuss its critical points.

\section{Critical points of the scalar potential}

According to our results in sect.~3, the critical points of the hypermultiplet
scalar potential are given by {\it poles} of $P$ and $w$, if one assumes 
that those poles are isolated points. This is the case, as long as the 
scalar potential is controlled by a meromorphic function like the Eisenstein 
series. Those poles precisely correspond to the points where the gauged 
$U(1)$ Killing vector (3.9) vanishes, because of eq.~(3.10):
$$ K^2=0\quad {\rm is~equivalent~to}\quad w^2P=\infty~,\eqno(5.1)$$
in agreement with the general results of refs.~\cite{clg,toy,fer}.

Generally speaking, eq.~(5.1) defines a (null) surface in the hypermultiplet 
moduli space, either of real dimension zero or two, depending upon the rank of
 the two-form $dK$ on the surface \cite{gh2}. If the rank is maximal, the null
 surface is just a point called {\it nut}. When the rank of $dK$ is two, the 
null two-dimensional surface is called a {\it bolt}. 

A physical vacuum is supposed to allow a perturbative expansion around it, 
which amounts to analyticity of the Killing vector and a finite curvature at 
the critical point, in our situation. The good (physical) critical points are 
therefore described by the following (NLSM) reparametrization-invariant 
conditions \cite{toy}:
$$ g_{XY}K^XK^Y\equiv K^2=0~,\quad (D_XK_Y)(D^XK^Y)\equiv (DK)^2\neq 0~,
\eqno(5.2)$$
and
$$R_{XYZW}R^{XYZW}\neq \infty~.\eqno(5.3)$$
 
The critical points of the scalar potential in the particular N=2 gauged 
supergravity model of a hypermultiplet, based on the non-homogeneous 
Einstein-Weyl metric interpolating between two homogeneous quaternionic 
metrics of $SO(4,1)/SO(4)$ and $SU(2,1)/U(2)$, were analyzed in detail by
Behrndt and Dall'Agata \cite{toy}. They found two good IR fixed points in 
their model \cite{toy}. Unfortunately, the hypermultiplet moduli space, used 
as an input in ref.~\cite{toy}, has a singularity, while it is not 
geodesically complete. Quantum instanton corrections to the classical 
hypermultiplet moduli space metric are expected to result in a regular
and positive definite metric \cite{bbs}. Further progress in this direction 
apparently requires an explicit knowledge of exact (non-separable) solutions 
to the 3d Toda equation (3.6).
  
In the particular case of the D-instanton-corrected UH moduli space, 
controlled by the Eisenstein series $E_{3/2}$, the situation is much simpler. 
The $E_{3/2}$ has polynomial growth $(\sim \r^{3/2})$  at weak coupling 
$\r=+\infty$ that corresponds to the classical vacuum, while all the one-loop 
and D-instanton quantum corrections disappear in that limit. This is 
consistent with our equations in sect.~4. When $\r\to +\infty$, we have
$$ E_{3/2}\sim \r^{3/2}~,\quad G\sim \r^2~,\quad 
\fracmm{\pa E_{3/2}}{\pa\h}\to 0~,\quad {\rm and}\quad \cw\to const.\neq 0~,
\eqno(5.4)$$
so that $K^2=G^{-2}\cw^{-1}\sim \r^{-2}\to 0$ indeed. Hence, the classical
limit 
$$\r=+\infty \eqno(5.5)$$
corresponds to a fixed point of the D-instanton-induced scalar potential.

As regards finite values of $\r\neq 0$ (i.e. strong coupling), the 
Eisenstein series $E_{3/2}$ has no singularities, while the function 
$G=\sqrt{\r}E_{3/2}$ is finite even at $\r=0$. Hence, any other critical 
points are only possible when $P=\infty$, i.e.
$$ \fracmm{\pa E_{3/2}}{\pa\r}+\fracmm{E_{3/2}}{2\r}=
\fracmm{\pa E_{3/2}}{\pa\h}=0~,\eqno(5.6)$$
where we have used eq.~(4.13). Equation (5.6) has a solution, 
$\r=0$ and $\h\in {\bf Z}$, where we have taken into account that the 
Eisenstein series is periodic in $\h$ with period $1$. Indeed, by using the 
relation $K_1(z)\approx z^{-1}$ for small values of $z\to 0$, it is not 
difficult to verify that for small values of $\r\to 0$ (at strong coupling), 
we have
$$ 4\p\z(3)E_{3/2}(\r,\h)\approx \r^{-1/2}\left[ \fracmm{2\p^2}{3}
+4\sum_{m\neq 0}\s_{-2}(m)e^{2\p im\h}\right]~,\eqno(5.7)$$
and
$$\fracmm{\pa G}{\pa\h}=\sqrt{\r}\fracmm{\pa E_{3/2}}{\pa\h}
=\fracmm{-4}{\z(3)}\sum^{+\infty}_{m=1}m\s_{-2}(m)\sin(2\p m\h)~,\eqno(5.8)$$
where we have introduced the standard divisor function \cite{ter}
$$ \s_s(m)=\sum_{0<d|m}d^s~,\eqno(5.9)$$
It is now clear that the only solution to eq.~(5.6) is given by
$$\r=0\quad {\rm and}\quad \h\in {\bf Z}~.\eqno(5.10)$$ 

We conclude that the D-instanton-induced scalar potential of the universal
hypermultiplet has the classical fixed point (5.5) at weak coupling and the 
fixed points (5.10) at strong coupling. The latter exactly appear at the 
points where the D-instantons are located, so that they are truly generated 
by the D-instantons.

\newpage

\section{Conclusion}

Our results may have natural applications to the brane-world scenarios 
\cite{bworld} within the effective N=2 supergravity originating from the 
CY compactified type-II superstrings and M-theory (see e.g., 
refs.~\cite{prob,bsve,stelle}). They may also be applied to a description of 
possible renormalization group flows in the holographic approach to extended 
supergravity \cite{hol}. The standard five-dimensional spacetime metric 
respecting the four-dimensional Poincar\'e invariance, is given by
$$ ds^2_{5d} = e^{2U(r)}dx^2_{4d}+dr^2~,\eqno(6.1)$$
where $U(r)$ is the warp factor \cite{bworld}. The domain wall solutions 
in the gauged five-dimensional N=2 supergravity normally preserve half of 
the original supersymmetries. In the case of a single hypermultiplet 
supporting the domain wall, the BPS (flow) equations are given by \cite{bsve}
$$ \fracmm{dU}{d\t}=\pm gW, \qquad \fracmm{dq^X}{d\t}=\mp 3gg^{XY}\pa_YW~,
\eqno(6.2)$$
where $W$ is the superpotential and $\t$ is the flow parameter. The equations 
of motion are automatically satisfied for the BPS solutions to eq.~(6.2). It
would be interesting to investigate the BPS solutions to eq.~(6.2) in the case
of the instanton-generated scalar potential. The BPS walls in some N=2 
supersymmetric non-linear sigma-models with hyper-K\"ahler metrics (in the 
absence of N=2 supergravity) were investigated in great detail in 
ref.~\cite{titech}.

The D-instanton corrections are given by powers of $e^{-1/g_{\rm string}}$, 
whereas the five-brane instantons contribute by powers of 
$e^{-1/g^2_{\rm string}}$ \cite{w1}. Our results apply when the former dominate
over the latter, i.e. when $g_{\rm string}$ is sufficiently small. The exact
quantum moduli space metric of UH is still governed by the same Toda equation 
(3.6), however, its general solution is unknown (see, however, ref.~\cite{nbi} 
for some explicit results about the five-brane instanton corrections to the UH
metric).

\section*{Acknowledgements}

Useful discussions with Ignatios Antoniadis and Norisuke Sakai are greatfully 
acknowledged. The author would like to thank the Yukawa Institute for 
Theoretical Physics in Kyoto for kind hospitality extended to him during a
preparation of this paper.

\newpage

\section*{Appendix: Einstein-Weyl geometry}

Our definitions and notation about the Einstein-Weyl spaces coincide with 
those used in refs.~\cite{prz,tod,iva}; see also ref.~\cite{besse} for more 
about the quaternionic geometry, and ref.~\cite{book} for more about the NLSM 
with quaternionic geometry and extended supersymmetry. We follow 
ref.~\cite{iva} here.

Given a four-dimensional Riemannian manifold of Euclidean signature with local
 coordinates $q^X$, 
$X=0,1,2,3$, and a metric $g=g\low{XY}dq^Xdq^Y$, let's introduce a local
basis (or a {\it vierbein}) $e\low{a}=e\low{aX}dq^X$ so that 
$$ g\low{XY}=\d\low{ab}e\low{aX}e\low{bY}~,\quad {\rm or}\quad 
g = \sum_a e^2_a~.\eqno(A.1)$$
The `time' $0$-direction is associated with an abelian isometry of the metric 
in the main text of the paper. We use capital Latin letters for curved 
4-vector indices and early lower-case Latin letters for flat (tangent) 
4-vector indices, whereas middle lower-case Latin indices denote `spatial' 
components of flat (tangent) 4-vector indices, $a=(0,k)$, $k=1,2,3$, in the
 hypermultiplet moduli space (NLSM).

The spin connection (1-form) $\o\low{ab}=\o\low{abX}dq^{X}$ is fixed by the
vierbein postulate, which means the covariant constancy of the vierbein, 
$$ de_a +\o_{ab}\wedge e_b=0~.\eqno(A.2)$$
The spin-connection is antisymmetric, $\o_{ab}=-\o_{ba}$, while
it can be decomposed into the {\it Self-Dual} (SD) and {\it Anti-Self-Dual} 
(ASD) parts, 
$${\o\low{ab}}^{\pm}=\o\low{ab}\pm\fracm{1}{2}\ve\low{abcd}\,\o\low{cd}~,\quad{\rm 
or,~equivalently,}\quad
{\o\low{i}}^{\pm}= \o\low{0i}\pm \fracm{1}{2}\ve\low{ijk}\o\low{jk}~.\eqno(A.3)$$

The curvature 2-form is defined by
$$ R_{ab} = d\o_{ab}+\o_{ad}\wedge\o_{db}=\frac{1}{2}R_{ab,cd}\, e_c\wedge e_d~,
\eqno(A.4)$$
while its SD and ASD components are given by
$$ {R\low{i}}^{\pm} = R\low{0i}\pm \fracm{1}{2}\ve\low{ijk}R\low{jk}~.\eqno(A.5)$$
The Ricci tensor and the scalar curvature are defined by
$$ (Ric)_{ab}=R_{ac,bc}~,\quad {\rm and}\quad R=(Ric)_{aa}~.\eqno(A.6)$$

The Weyl curvature is given by the traceless part of the curvature, {\it viz.}
$$\eqalign{
W_{ab,cd} ~=~& R_{ab,cd}+\fracmm{R}{6}\left[ \d_{ac}\d_{bd}-\d_{ad}\d_{bc}
\right]\cr
 ~& -\fracmm{1}{2}\left[ \d_{ac}(Ric)_{bd}-\d_{ad}(Ric)_{bc}
+\d_{bd}(Ric)_{ac}-\d_{bc}(Ric)_{ad}\right]~.\cr}\eqno(A.7)$$

The 2-forms $e_a\wedge e_b$ serve as a basis in the space of all 2-forms, while 
they can also be decomposed into their SD and ASD parts,
$$ \X^{\pm}_i= e_{0}\wedge e_{i} \pm \frac{1}{2}\ve_{ijk}e_{j}\wedge e_{k}~.\eqno(A.8)$$
In particular, we have
$$ \eqalign{
R^+_i & =  A_{ij}\X^+_j + B_{ij}\X^-_j~,\cr
R^-_i & =  B^T_{ij}\X^+_j + C_{ij}\X^-_j~,\cr}\eqno(A.9)$$
where the symmetric $3\times 3$ real matrices $A$ and $C$, and the 
non-symmetric $3\times 3$ real matrix $B$ have been introduced. 

Similarly, the Weyl 2-form  
$$W_{ab}=\frac{1}{2}W_{ab,cd}\, e_c\wedge e_d\eqno(A.10)$$ 
can be decomposed into its SD and ASD parts as
$$ \eqalign{
W^+_i = & W_{0i} + \frac{1}{2}\ve_{ijk}W_{jk} = W^+_{ij}\X^+_j~,\cr
W^-_i = & W_{0i} - \frac{1}{2}\ve_{ijk}W_{jk} = W^-_{ij}\X^-_j~.\cr}
\eqno(A.11)$$

The {\it Einstein} condition (with a real constant $\L$),
$$ (Ric)_{ab} =\L \d_{ab}~,\eqno(A.12)$$
is equivalent to
$$ B_{ij}=0~ \quad {\rm and} \quad \tr\,A =\tr\,C=\L~.\eqno(A.13)$$

The self-duality of the {\it Weyl\/} tensor,
$$ W_i^-=0~,\eqno(A.14)$$
is equivalent to
$$ C_{ij}\propto \d_{ij}~.\eqno(A.15)$$

Hence, the SD Weyl and Einstein conditions together imply 
$$ C_{ij} =\fracmm{\L}{3}\,\d_{ij}\quad {\rm and}\quad R^-_i=
\fracmm{\L}{3}\,\X^-_i~.\eqno(A.16)$$
The only remaining matrix $A$ is symmetric and has $\tr\, A =\L$, so that
there are five independent curvature components for a generic Einstein-Weyl
metric.


\newpage

\begin{thebibliography}{99}
\bibitem{bbs} K. Becker, M. Becker and A. Strominger, \np{456}{95}{130} 
\newline [hep-th/9507158] 
\bibitem{fsh} S. Ferrara and S. Sabharwal, \cqg{6}{89}{L77}; \np{332}{90}{317}
\bibitem{one} P. Candelas, X.C. de la Ossa, P.S. Green and L. Parkes,
\np{359}{91}{21};\\
A. Strominger, \pl{421}{98}{139} [hep-th/9706195] 
\bibitem{w1} E. Witten, \np{474}{96}{343} [hep-th/9604030]
\bibitem{actions} K. Becker and M. Becker, \np{551}{99}{102} 
[hep-th/9901126];\\
M. Gutperle and M. Spalinski, JHEP {\bf 0006} (2000) 037  
 [hep-th/0005068], and  Nucl. Phys. {\bf B598} (2000) 509 [hep-th/0010192];\\
U. Theis and S. Vandoren, JHEP {\bf 0209} (2002) 059 [hep-th/0208145]
\bibitem{bw} J. Bagger and E. Witten, \np{222}{83}{1}
\bibitem{ket} S. V. Ketov, {\it D-Instantons and universal hypermultiplet},
CITUSC preprint 01--046 [hep-th/0112012], unpublished; Nucl. Phys. {\bf B649}
 (2003) 365 [hep-th/0209003]
\bibitem{green} M. B. Green and M. Gutperle, Nucl. Phys. {\bf B498} (1997) 195
 [hep-th/9701093];\\
M. B. Green, M. Gutperle and P. Vanhove, Phys. Lett. {\bf 408B} 
(1997) 122 \newline [hep-th/9704145], and Phys. Lett. {\bf 409B} (1997) 177 
 [hep-th/9706175]
\bibitem{nbi} S. V. Ketov, Phys. Lett. {\bf 504B} (2001) 262 
[hep-th/0010255]; Nucl. Phys. {\bf B604} (2001) 256 [hep-th/0102099]
\bibitem{pen} S. V. Ketov, Fortschr. Phys. {\bf 50} (2002) 909 [hep-th/0111080]
\bibitem{ps} J. Polchinski and A.  Strominger, Phys. Lett. {\bf 388B} 
(1996) 736 [hep-th/9510227];\\
J. Michelson, Nucl. Phys. {\bf B495} (1997) 127
 [hep-th/9610151];\\
T. R. Taylor and C. Vafa, Phys. Lett. {\bf 474B} (2000) 130  
[hep-th/9912152];\\
G. Dall'Agata, JHEP 0111 (2002) 005 [hep-th/0107264]
\bibitem{clg} K. Behrndt and M. Cveti\v{c}, Nucl. Phys. {\bf B609} (2001) 183
[hep-th/0101007];\\
A. Ceresole, G. Dall'Agata, R. Kallosh and A. Van Proeyen, 
Phys. Rev. {\bf D64} (2001):104006 [hep-th/0104056]
\bibitem{bworld} V. A. Rubakov and M. E. Shaposhnikov, \pl{125}{83}{136};\\
 L. Randall and A. Sundrum, Phys. Rev. Lett. {\bf 83} (1999) 4690
\bibitem{prob} R. Kallosh and A. Linde, JHEP {\bf 02} (2000) 005
[hep-th/0001071];\\
 K. Behrndt and M. Cveti\v{c}, Phys. Rev. {\bf D61} (2000) 101901 
[hep-th/0001159]
\bibitem{toy} K. Behrndt and D. Dall'Agata, Nucl.Phys {\bf B627} (2002) 357 
[hep-th/0112136]
\bibitem{sbl} L. Anguelova and C. I. Lazaroiu, JHEP {\bf 0209} (2002) 053
  [hep-th/0208154] 
\bibitem{fer} R. D'Auria and S. Ferrara, JHEP {\bf 05} (2001) 034 
[hep-th/0103153]
\bibitem{prz} M. Przanowski, J. Math. Phys. {\bf 32} (1991) 1004
\bibitem{tod} K. P. Tod, Twistor Newsletter {\bf 39} (1995) 19; 
Lecture Notes in Pure and Appl. Math. {\bf 184} (1997) 307 
\bibitem{iva} P.-Y. Casteill, E. Ivanov and G. Valent, Nucl. Phys. {\bf B627} 
 (2002) 403 \newline
[hep-th/0110280] 
\bibitem{ward} M. Saveliev, Commun. Math. Phys. {\bf 121} (1989) 283;\\
R. S. Ward, Class. and Quant. Grav. {\bf 7} (1990) L95
\bibitem{lez} A. N. Leznov, Theor. Math. Phys. {\bf 117} (1998) 1194
\bibitem{lebrun} C. R. Lebrun, J. Diff. Geom. {\bf 34} (1991) 223
\bibitem{gh} G. W. Gibbons and S. W. Hawking, Phys. Lett. {\bf 78B} (1978) 430
\bibitem{tong} J. P. Gauntlett, D. Tong and P. K. Townsend, Phys. Rev. 
{\bf D63} (2001) 085001 [hep-th/0007124], and Phys. Rev. {\bf D64} (2001) 
025010 [hep-th/0012178]]
\bibitem{cp} D. M. J. Calderbank and H. Pedersen, J. Diff. Geom. {\bf 60}
 (2002) 485  math.DG/0105263;\\
D. Calderbank and M. A. Singer, {\it Einstein metrics and
complex singularities},  math.DG/0206299 
\bibitem{ter} A. Terras, {\it Harmonic Analysis on Symmetric Spaces and
Applications}, Springer-Verlag, 1985, p.p. 204--228
\bibitem{gh2} G. W. Gibbons and S. W. Hawking, \cmp{66}{79}{291}
\bibitem{bsve} K. Behrndt and M. Cveti\v{c}, Phys. Let. {\bf 475B} (2000) 253 
[hep-th/9909058];\\
K. Skenderis and P. K. Townsend, \pl{468}{99}{46} [hep-th/9909070]
\bibitem{stelle} A. Lukas, B. A. Ovrut, K. S. Stelle and D. Waldram, Phys. Rev.
{\bf D59} (1999) 086001 [hep-th/9803235]
\bibitem{hol} E. Alvarez and C. Gomez, \np{541}{99}{441} [hep-th/9807226];\\
L. Girardello, M. Petrini, M. Porrati and A. Zaffaroni, JHEP {\bf 12} (1998) 
022 \newline
[hep-th/9810126];\\
D. Z. Freedman, S. S. Gubser, K. Pilch and N. P. Warner, Adv. Theor. Math. 
Phys. {\bf 3} (1999) 363 [hep-th/9904017];\\
S. V. Ketov, Nucl. Phys. {\bf B597} (2001) 245 [hep-th/0009197] 
\bibitem{titech} M. Naganuma, M. Nitta and N. Sakai, Grav. and Cosmology 
{\bf 8} (2002) 129 \newline [hep-th/0108133];\\
M. Arai, M. Naganuma, M. Nitta and N. Sakai, {\it Manifest supersymmetry for
BPS walls in N=2 non-linear sigma-models}, Tokyo preprint TIT/HEP--487
\newline [hep-th/0211103];\\
N. Sakai, {\it Supersymmetry}, Lectures given by. N. Sakai at TMU in 
November 2002, unpublished
\bibitem{besse} A. L. Besse, {\it Einstein Manifolds}, Springer-Verlag, 1987
\bibitem{book} S. V. Ketov, {\it Quantum Non-linear Sigma-Models}, 
Springer-Verlag, 2000.

\end{thebibliography}
\end{document}